\documentclass[preprint,12pt]{elsarticle}

\usepackage{lineno,hyperref}

\journal{Nuclear Instruments and Methods}  

\begin{document}

\begin{frontmatter}

\title{The ECFA Roadmap Process for Particle Identification   and Photon Detector R\&D
}

\author[address1]{N. Harnew}
\fntext[myfootnote0] {Proceedings of the XI International Workshop on Ring Imaging Cherenkov Detectors}
\ead{neville.harnew@physics.ox.ac.uk}
\address[address1]{Denys Wilkinson Laboratory, University of Oxford, Keble Road, Oxford OX1 3RH, UK}

\begin{abstract}
The Detector R\&D Roadmap for  European Particle Physics was published in February 2022. The  outcome of the Roadmap process relating to particle identification   and photon detectors is  summarised.
\end{abstract}

\begin{keyword}
\texttt{ Particle identification \sep RICH detectors \sep Time-of-flight \sep  Photon detectors }
\MSC[2010] 00-01\sep  99-00
\end{keyword}

\end{frontmatter}


\section{The  European particle physics R\&D strategy }

For many decades, particle identification (PID) methods have been an indispensable experimental tool in elementary particle  physics, in particular for heavy flavour physics, heavy-ion collisions, and electron and hadron experiments,  and also in particle astrophysics.
The ever-growing demands of the future physics programme, from underground facilities to high-luminosity colliders, requires a new generation of PID detectors with separation power over 4-5 orders of magnitude in momentum.
In addition, advancement in photon detector technology is essential to address not only PID detectors, but also  the science requirements of all future high energy physics experiments. Photon measurements are paramount in   calorimetry, tracking, neutrino and dark-matter experiments, covering applications of ultra-high rates to extreme low-noise requirements.

Organised by ECFA, a Roadmap has been developed to balance the detector R\&D efforts in Europe, taking into account progress with emerging technologies in adjacent fields~\cite{ECFARoadMap}.  
The Roadmap has identified  a diversified detector R\&D portfolio with the aim to enhance the performance of the particle physics programme in the near and long-term future.
The Roadmap identifies detector R\&D activities that require specialised infrastructures, tools, and access to test facilities.
It defines activities that can be used to support research proposals at the national and international  levels.

The  European R\&D strategy process was launched by the CERN Council in September 2018 and a kick-off meeting was held on 21-22 Feb 2021 comprising twelve talks which discussed the requirements of the future facilities listed below~\cite{ECFA-facilities}.

\begin{itemize}
\item  $<$2030 :Storage rings and fixed target  (Belle-II, Rare kaon, TauFV, Mu3e,  etc),  
\item $<$2030 : Neutrino Long Baseline,  
\item  $<$2030  : Nuclear physics applications (GSI, Compass2),   
\item 2030--2035 : HL-LHC (ATLAS/ CMS/ LHCb),
\item 2030--2045 : High-energy electron-hadron/ion  collider (EIC, ALICE3,  FCC-eh/eA),  
\item 2035--2045 :Higgs-Top Factories  (FCC-ee, ILC),   
\item $>$2045 : Muon collider, FCC-hh. 
\end{itemize}

\begin{figure}[tb]
  \begin{center}
    \includegraphics[width=1.0\linewidth]{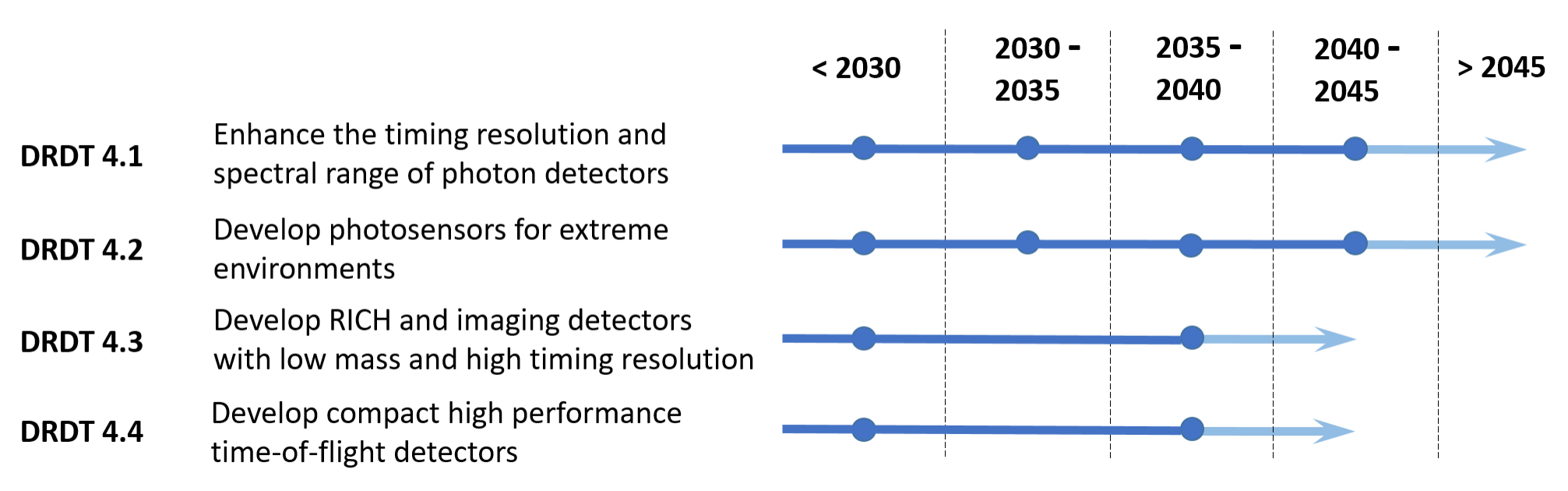} 
  \end{center}
  \caption{
Detector themes  for  photon detectors and PID (DRD4). The dots on the time-lines represent the milestones which result from  the  facility's requirements.
    }
  \label{fig:DRDT}
\end{figure}

A total of nine task forces (TFs) were established to define strategy in the relevant areas, TF4 being assigned to PID and photon detection. A remote symposium for TF4 was held on 6 May 2021 for community consultation and discussion, which had  110    attendees~\cite{ECFA-TF4-symposium}. 
Each task force defined a number of Detector Research and Development Themes (DRDTs) 
with corresponding time-lines. Those related to photon detectors and PID, namely DRD4,  are shown in Fig.~\ref{fig:DRDT}. 

The four major requirements specified for DRD4 are listed below.  
\begin{itemize}
\item DRD 4.1 : Enhance the timing resolution and spectral range of photon detectors. \\
This relates to the development of fast timing for Cerenkov and time-of-flight detectors, operation with high particle fluxes and pile-up, and in extending the wavelength coverage of scintillation photons from noble gases and Cerenkov photons.
\item DRDT 4.2 : Develop photosensors for extreme environments. \\
This is essential for operation in the high-radiation environments at the HL-LHC, the Belle II Upgrade, EIC and FCC-hh,  and similarly for cryogenic operation.
\item DRDT 4.3 :  Develop RICH and imaging detectors with low mass and high resolution. \\
This is required for particle identification at the HL-LHC, the Belle II Upgrade, EIC, and FCC-ee.
\item DRDT 4.4 : Develop compact high performance time-of-flight detectors. \\
This is  as a complementary approach for particle identification at HL-LHC, EIC and FCC-ee.
\end{itemize}

\section{RICH, DIRC and TOP  detector technologies}

Ring Imaging Cherenkov (RICH) detectors~\cite{Seguinot:1976yp} have evolved to unprecedented levels of performance~\cite{KRIZAN2017272}. This has allowed very efficient PID of charged particles coupled with   outstanding background rejection over a vast momentum range, from a few 100\,MeV/c up to several 100\,GeV/c. Over the last three decades, there has been the evolution of Detection of Internally Reflected Cherenkov light (DIRC) detectors~\cite{Adam:2004fq} and more recently the Time of Propagation (TOP) detector~\cite{2014NIMPAKenji} which measures the time of arrival of the Cherenkov photons. Despite the huge success of these detectors, further innovations of the technologies are required to meet the needs of the future facilities.

\subsection {Enhancement  of  RICH technologies} 
The HL-LHC has possibly the most demanding requirements for the next generation of RICH detectors where the  momentum coverage is between 10 to 200\,GeV/c .The R\&D focus is photon detection  in the visible wavelength region with a reduction in chromatic dispersion, with an aspiration of 
a  0.3--0.5\,mrad single-photon angular resolution.  
Hermetic experiments (eg FCC-ee, EIC) need RICH detectors with a total length shorter than a meter. This introduces the 
possibility of pressurizing noble gases to several bars to increase the refractive index which in turn requires innovative engineering solutions. 
Design of compact RICH detectors working in the far-UV region to improve photon yield are also a consideration. Smart optics design is  important: 
improvements in lightweight mirror technology (in particular with cost-effective carbon-fibre), improved light collection systems and concentrators are necessary.

There needs to be extensive R\&D into new radiator materials.
So-called green gas radiators  need to be developed  as  alternatives to the traditional  fluorocarbon gases (C$_4$F$_{10}$ or CF$_4$).
Aerogel radiators with tunable refractive indices, larger tiles, and higher photon yield need to be investigated, together with the availability of more production sites.  The technique of dual radiators  should be extended to future facilities.
The R\&D into meta-materials and photonic crystals for RICH radiators  is very much in its infancy, however the possibility exists to match the refractive indices needed to identify particles at low and high momenta.
Neutrino experiments  for large volumes have different requirements, where high optical purity of liquid radiators  needs to be improved.

Photon detectors are a key technology in the evolution of future RICH detectors.
Large-area single-photon detectors capable of sustaining high counting rates (a few MHz/cm$^2$) with a total ionizing dose up to a few Mrad are required, with improved photon detection efficiency and timing resolution of the order of few tens of ps per photon. Granularities are required of a mm$^2$-level pixel size.
Photon detectors of choice are silicon photomultipliers (SiPMs),  microchannel-plate photomultipliers (MCP-PMTs) and solid photocathodes in micro-pattern gaseous detectors (MPGDs).

Regarding readout electronics, timing of photon arrival will be absolutely crucial to the new generation of RICHes (with $<$50\,ps precision) in order to improve background rejection and  association of collision vertices due to multiple interactions (pile-up). ASICs need to be radiation-hard, have a low power consumption, have a few tens of picosecond resolution, and high-granularity to satisfy photon angular-resolution requirements.

\subsection {DIRC and TOP detectors}
The advantage of a DIRC detector is that it provides an extremely compact PID device, however current detectors are suitable only for low momentum PID up to 4--5\,GeV/c. 
Applications are for flavour and heavy-ion physics, where PID is essential. e.g. Panda, Belle-II, Gluex and later EIC and HL-LHC; it is the latter two which drive the performance requirements.  The future aim is to extend to 3 sigma $K/\pi$ separation up to at least 7--8\,GeV/c (with TOP correction).

R\&D is required to make DIRC geometry more compact, expand the momentum reach, and further development for endcap operation. Improvements in focusing designs are needed, with emphasis on the spatial resolution.
Multiple scattering should be reduced and other RICH resolution terms should be mitigated, including chromatic dispersion.  
Highly polished quartz radiator bars are essential, and it is necessary to develop quartz technology at much cheaper cost. Here the surface quality is paramount  and (sub)-nm surface roughness is required.

For photon detection,  state of the art timing is essential  ($\mathcal{O}$(10) ps binning). Devices of choice are  SiPMs and  MCP-PMTs   (the latter at   reduced cost) with improved and customised photon-detector granularity. Radiation tolerance, long lifetime (for years of high-luminosity running), low noise and good photon sensitivity are prerequisites.

\subsection {Time-of-flight technologies }
Time-of-flight (TOF) discrimation falls fast with velocity, and benefits from a long flight distance.  The technique  therefore provides only complimentary PID al low momentum for future high-energy machines. Nevertheless, the method is an essential tool for experiments such as  Panda/CBM, Na62/TauLV, and ALICE and is being considered for future upgrades of LHCb, ATLAS/CMS, and at the EIC and FCC.

There are various TOF methods which need to undergo future R\&D~\cite{Cavallari_2020}. The enhancement of scintillator timing down to a resolution of $\sim$50\,ps, gaseous detectors such as   micro-pattern gas detectors (MPGDs) and multigap RPCs to $<$50\,ps. 
Silicon sensors such as low gain avalanche diodes (LGADs) will be important for the LHC General Purpose Detectors  and FCC-hh, where timing will be important to $\mathcal{O}$(10)\,ps. 
Another technology that needs to be further developed is the large-area MCP-PMT with timing down to 15\,ps. Here Large Area Picosecond Photodetectors (LAPPDs) can in principle instrument  large  surface areas~\cite{Minot:2020hlb},  but necessarily at reduced cost. 
Cherenkov (DIRC)-based TOF detectors such as the TORCH detector~\cite{BHASIN2023168181} for LHCb, TauFV and kaon physics can provide 3 sigma $K\pi$ separation to 10\,GeV/c over a 10\,m flight path. Here MCP-PMTs with high granularity (32 $\times$ 128 pixels over a 2-inch$^2$ active  area) must  withstand integrated charges of $\sim$50-100\,Ccm$^{-2 }$. Alternaively SiPMs are  an option which must be investigated. Highly-polished quartz technology must also be developed at cheaper cost, as for the DIRC. 

For all TOF applications, a state-of-the-art timing ASIC will be compulsory ($\mathcal{O}$(10) ps binning)  with low noise amplification/ discrimination and a high channel count. 
R\&D will be essential for timing (clock) distribution across several modules, together with a well-defined $t_0$.  

\section{Photon detector technologies}

Photon detection remains a key ingredient of the majority of  PID devices, as well as having essential applications in calorimetry and, to a lesser extent, tracking.
Several vacuum-based  photon detector technologies have evolved from the classic photomultiplier concept over the past decades, and still have wide applications for PID~\cite{Gys:2020edc}. SiPMs are photosensors of choice for many applications and are therefore  a key technology~\cite{Gundacker_2020}; here the HL-LHC and FCC-hh mainly drive the state-of-the-art  requirements.

\subsection {Family of vacuum-based photodetectors} 

MCP-PMTs have been proposed extensively for RICH detectors and TOF, 
where the state-of-the-art  timing resolution is currently around $\sim$20\,ps~\cite{Bohm:2020njp}.
There are several suppliers  e.g. Incom (large area),  Photonis,  Hamamatsu and Photek (who have produced compact, customised-granularity devices).
There are several significant challenges that require R\&D to make the devices suitable for the next-generation experiments. 
Large area coverage has been addressed above by LAPPDs, but cheaper cost and mass production has to be demonstrated for tiling large areas.
Customised  granularity for DIRC-type detectors and TOF needs to be convincingly demonstrated;  approaching \mbox{128 $\times$ 128} pixels over a 2-inch$^2$  tube size would likely be the practical limit, but then with suitable electronics grouping to reduce the overall channel count. 
The rate limitation of MCP-PMTs needs to be addressed, currently around  10$^5$/cm$^2$,   with an improvement necessary to beyond a  MHz.
The integrated charge capability also needs to be improved by several factors above the current state-of-the-art of $\sim$20 C/cm$^2$.
This can be achieved by reducing the nominal operating gain by at least an order of magnitude, currently 10$^6$--10$^7$, but low-noise  electronics would need to facilitate operation at a few $\times$10$^5$ to be effective.
Improvements in quantum efficiency (QE) and charge collection efficiency (CCE) also need to be achieved. 

Standard photomultipliers (including large area PMTs) are in demand for neutrino and non-accelerator physics applications. Here
improvements are needed to increase the radio-purity by a factor of 5--10, the UV response and the QE.
Other devices less in current demand but with possible useful applications are  Multianode PMTs (MaPMTs),  hybrid avalanche photon detectors (HAPDs), hybrid HPDs (which already have an active R\&D programme) and vacuum silicon photomultiplier tubes (VSiPMTs).

\subsection{Gas-based photodetectors}
Gaseous Photon Detectors (GPDs) represent an effective solution for instrumenting large imaging surfaces (up to several square metres) in high magnetic felds~\cite{sauli_2014}.
There is a requirement  to develop GPDs based on Micro-Pattern Gaseous Detector  (MPGD) structures which allow photon conversions at the level of a  few tens of microns~\cite{OED1988351}.
Further R\&D  is necessary to improve the photocathode lifetime (GHz rates are required for the EIC), the photon detection efficiency (PDE), radiation hardness and the time resolution in the few ps range. 
There is also a need to develop compact GPD systems with integrated electronics for imaging applications. 

UV-sensitive materials which are more radiation-hard and chemically inert than CsI will be required.
A further challenge will be the extension to the visible spectral range; carbon-based photocathodes need to be studied. 
Finally, R\&D for alternative hydrocarbon-free gas mixtures, 
extension to cryogenic applications, 
and detection of both scintillator light and ionization is necessary.

\subsection{Silicon photomultipliers}

Important features of SiPMs are their compactness, low operation voltage, robustness to magnetic fields and relatively low cost.
As well as for PID, they have wide applications for scintillating fibres, calorimeters, neutrinos and dark matter experiments, noble liquid detectors  
and  gamma ray astronomy~\cite{Simon_2019}. 
SiPMs are now becoming the detector of choice for RICH and DIRC-type detectors for the LHCb and ALICE experiments, and also for the EIC and FCC-ee. 
Their advantages are well known: SiPMs have a QE which  is typically around 50--60\% in the visible (350--600\,nm), they have fast timing response (significantly below 100\,ps), and  small cell sizes which are in principle tuneable.
However, disadvantages include their high dark count rates (10--100\,kHz mm$^{-2}$ at room temperature), hence the need for cooling. 
Their radiation hardness also needs significant improvement since for PID applications they lose sensitivity to single photons at around 1$\times$10$^{11}$ neutrons cm$^{-2}$ equivalent (n cm$^{-2}$ eq.).

There is an extensive list of SiPM  R\&D technology demands due to several diverse and sometimes conflicting applications. 
For HEP applications, the ATLAS and CMS upgrades and the FCC-hh drive the requirements.
Timing characteristics  need improvement,  with an aspiration to  achieve a  time resolution of $\sim$10\,ps.   
The radiation hardness is a key R\&D item with requirements of 1$\times$10$^{14}$ n cm$^{-2}$ eq. at ATLAS/CMS and 10$^{17}$--10$^{18}$ at FCC-hh.
Improved dark count rates towards 1\,Hz mm$^{-2}$ is necessary, especially important for low-light-level experiments where pulse-shape discrimination is also an advantage, resulting in reduced after-pulsing. Optimised  cooling systems at $-$50$^{\circ}$C  to reduce dark noise is required  for all future collider experiments.
Relevant to RICH and TOF detectors, there is a need to increase the PDE  for single-photons, increasing the fill factor, and extending the spectral range into the UV and infrared. There needs to be customised cell size, optimised optical couplings and development of micro-lenses/filters. Cheaper solutions for SiPMs (e.g. CMOS) for large area applications and high pixel density needs improvement, and also the development of  digital, as well as analogue,  readout.

\section{Summary of recommendations}
\begin{figure}
  \begin{center}
    \includegraphics[width=1.0\linewidth]{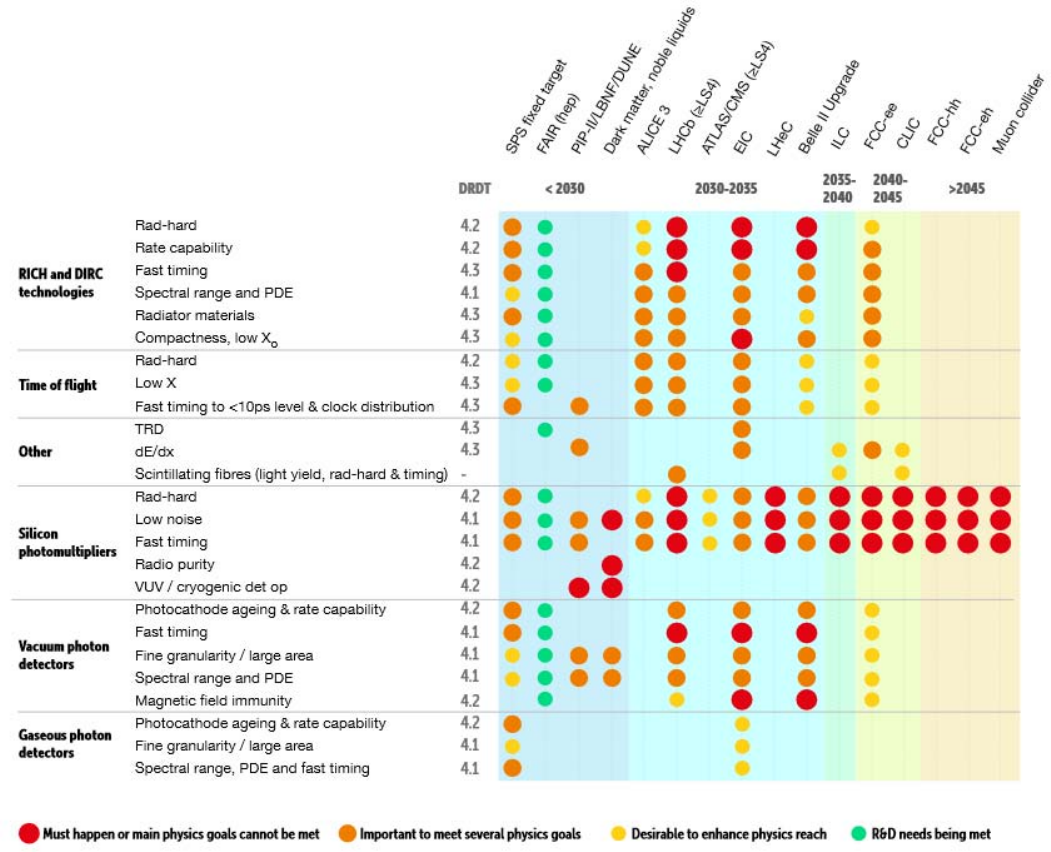} 
  \end{center}
  \caption{
Schematic time-line of categories of experiments employing PID and photon
detectors together with DRDTs and R\&D tasks. Red, largest dot - the technology must happen or the main physics goals cannot be met;
orange, large dot - the technology is important to meet several physics goals ;  yellow, medium dot - the technology is desirable to enhance the physics
reach; (green, small dot - the R\&D needs are already being met; blank - no further R\&D 
is required or not applicable. 
    }
  \label{DRD4profile}
\end{figure}

Figure\,\ref{DRD4profile} shows the time-line of categories of experiments employing PID and photon detectors, together with   the major DRD4 R\&D technology drivers.  
The colour coding is linked to the potential impact of the technique on the physics programme of
the experiment.

\begin{itemize} 

\item DRDT 4.1 : To enhance the timing resolution and spectral range of photon detectors.\\
Over the next five years, the community needs advances in SiPM technology  for fast timing, UV sensitivity and dark-count rates.
MCP-PMTs need  improved QE,  charge collection efficiencies, granularity and be more cost-effective for large areas. 
 Light collection systems in general require  improvements.
Over the next ten  years there needs to be incremental improvements to gaseous photon detector technology for granularity and fast timing.

\item DRDT 4.2 : Develop photosensors for extreme environments. \\
Over the next five years,  the radiation tolerance of SiPMs needs to be significantly enhanced.  
MCP-PMTs need improved detector ageing and high-rate performance.
The  photocathode ageing and rate capability of gaseous detectors needs to be improved.
Over the next 20 years there needs to be further advances in SiPMT radiation hardness to a couple of orders of magnitude beyond 1$\times$10$^{14}$ n cm$^{-2}$ eq.

\item DRDT 4.3 : Develop RICH and imaging detectors with low mass and high  resolution. \\
Over the next five years, there needs to be development of RICH  detectors with picosecond timing, greenhouse-friendly radiator gases and transparent aerogels,  and  cheaper quartz for DIRCs.
Over the next 10 years,  compact RICH systems with low radiation length  (e.g. pressurised systems) need to be progressed.

\item DRDT 4.4 : Develop compact high performance time-of-flight detectors. \\
Over the next five years, TOF systems with picosecond timing,  equipped with  high-granularity photosensors with long lifetime and high-rate capabilities, need to be developed. 

\end{itemize}

So-called ``blue sky'' R\&D activities should not be forgotten. Examples are the study of 
solid-state photon detectors from novel materials, the development of cryogenic superconducting photosensors, the study of gaseous photon detectors sensitive to visible light, and
metamaterials to allow tuneable refractive indices.

Finally,  innovative photon sensors require development in specialized industrial companies and, to be commercially attractive, standardisation across research areas must be fostered. A close synergy needs to be developed, in particular with other fields, to transfer HEP expertise and vice versa. Close collaboration is essential to ensure supply and lower cost,  and here SiPMs are a good example.   
Conversely, PID detectors are in general developed within HEP research institutes. It is important to develop a sustainable plan to recruit and train future instrumentalists to avoid generational gaps and to maintain excellent technical facilities.

\section{Implementation of the Roadmap}
As preparation towards the implementation of the Roadmap recommendations, the R\&D Task Forces will organise open meetings to establish the scope and scale of the communities wishing to participate in the new DRD activities. DRD conveners and   teams of experts will then be  identified. 
By Autumn  2023, DRD proposals will be prepared with work-package structures. 
By early 2024 the new DRDs will come into existence and the  R\&D programmes will get underway.
Throughout 2024 there will be a collection of Memorandums of Understanding (MoUs), with defined areas of interest amongst institutes. 
Throughout 2024--2026 there will be a ramp-up of new strategic funding and R\&D activities will proceed in parallel towards completion of a series of specified  deliverables. The R\&D activities will then evolve in the  future to enable new innovative technologies to sustain the needs of high-energy physics over the coming decades.

\section*{Acknowledgments}
I would like to thank the organisers of  RICH2022  for making the conference so enjoyable; in particular the conference chair  Eugenio Nappi and the local team chaired by Franz Muheim. I would also like to thank my co-convener on TF4, Peter Kri$\check{\rm z}$an,  the task-force panel members Ichiro Adachi, Christian Joram,  Eugenio Nappi  and  Hans-Christian Schultz-Coulon, and  the organisers of the ECFA process, Philip  Allport  and  Felix Sefkow. All provided essential  wisdom and guidance  towards the successful development and publication of the  ECFA Roadmap. 

\section*{References}

\bibliography{Harnew-bibfile}

\end{document}